\documentclass[final,5p,authoryear,10pt]{elsarticle}
\pdfoutput=1
\usepackage{amsmath}
\usepackage{graphicx}
\usepackage{booktabs}

\usepackage{amssymb}
\usepackage{amsthm}
\usepackage{xfrac}
\usepackage{float}
\usepackage{hyperref}
\usepackage{fancyvrb}
\usepackage{color}
\usepackage{epstopdf}
\usepackage{footnote} 
\usepackage[usenames,dvipsnames,svgnames,table]{xcolor}
\usepackage{subfig}

\usepackage[british]{babel}
\newcommand{\ba}{\begin{eqnarray}}
\newcommand{\ea}{\end{eqnarray}}

\newcommand{\mb}[1]{\mathbf{#1}}
\newcommand{\bs}[1]{\boldsymbol{#1}}

\newcommand{\oliver}[1]{#1} 
\newcommand{\Figure}{\oliver{Figure}}

\journal{Annals of Nuclear Energy}

\begin{document}

\begin{frontmatter}



\title{Improving PWR core simulations by  \linebreak[9] Monte Carlo uncertainty analysis and Bayesian inference}

\author[UPM,AREVA]{E.~Castro}
\ead{emilio.castro@upm.es}
\author[UPM]{C.~Ahnert}
\author[AREVA]{O.~Buss}
\author[UPM]{N.~Garc\'{i}a-Herranz}
\author[AREVA]{A.~Hoefer}
\author[AREVA]{D.~Porsch}

\address[UPM]{Universidad Polit\'{e}cnica de Madrid - Dpto. de Ingenier\'{i}a Energ\'{e}tica, \'{a}rea de Ingenier\'{i}a Nuclear\\
	C. Jos\'{e} Guti\'{e}rrez Abascal 2, 28006 Madrid, Spain.}
\address[AREVA]{AREVA~GmbH, Paul-Gossen-Strasse 100, 91052 Erlangen, Germany}

\begin{abstract}
A Monte Carlo-based Bayesian inference model is applied to the prediction of reactor operation parameters of a PWR nuclear power plant. In this non-perturbative framework, high-dimensional covariance information describing the uncertainty of microscopic nuclear data is combined with measured reactor operation data in order to provide statistically sound, well founded uncertainty estimates of integral parameters, such as the boron letdown curve and the burnup-dependent reactor power distribution. The performance of this methodology is assessed in a blind test approach, where we use measurements of a given reactor cycle to improve the prediction of the subsequent cycle. As it turns out, the resulting improvement of the prediction quality is impressive. In particular, the prediction uncertainty of the boron letdown curve, which is of utmost importance for the planning of the reactor cycle length, can be reduced by one order of magnitude by including the boron concentration measurement information of the previous cycle in the analysis. Additionally, we present first results of non-perturbative nuclear-data updating and show that predictions obtained with the updated libraries are consistent with those induced by Bayesian inference applied directly to the integral observables.
\end{abstract}

\begin{keyword}
Uncertainty analysis \sep Nuclear data \sep Monte Carlo methods \sep PWR core analysis \sep Bayesian inference

\end{keyword}

\end{frontmatter}


\newpage

%
\section{Introduction}
\label{sect::intro}
%
%
%
Best estimate plus uncertainty methodologies allow for a reliable quantification of safety margins for nuclear power plant operation and for the manufacturing, handling, storage and transport of nuclear fuel \citep{margins_IAEA,be_IAEA,uam_vol1,uacsa_report}. This opens up the possibility to eliminate unnecessary conservatism, which leads to improved plant performance through optimized core design and greater operational flexibility and to reduced fuel management costs. However, the evaluation of statistical confidence bounds for nuclear safety parameters requires a consistent statistical framework to combine and propagate uncertainties in nuclear data, technological data, and operational data.

Two different approaches are currently used to propagate nuclear data uncertainties to integral observable uncertainties: perturbation theory and Monte Carlo sampling. 

The perturbation theory approach has been successfully used for more than a half a century \citep{uchasev,gandini} and has been implemented in many different computer codes. Within this framework, integral functions of nuclear data are approximated by their first order series expansions, which implies that the integral variable uncertainties are expressed as linear transformations of nuclear data covariances defined by the sensitivities of the integral observables to the nuclear data; see e.g.~\citep{broadhead}. Typically, the applied nuclear transport codes have to be upgraded for computing the sensitivity coefficients. As a first order approximation, this approach yields sufficiently accurate uncertainty estimates only under the condition that the relevant nuclear data uncertainties are not too large. 

In recent years, the Monte Carlo sampling approach has been playing an increasingly important role in the uncertainty propagation of nuclear data to integral observables \citep{koning_tmc, acab_2008, xsusa, nuduna,xsusa_scale, psi}. This method consists in random sampling of nuclear data parameters - optionally together with technological and operational parameters - from their joint uncertainty distribution, where each random sample is used in a different computation of the integral variable of interest. Finally, uncertainty estimations for the integral variables are obtained from the statistics of Monte Carlo computations. Since this approach does not rely on perturbation theory, any inaccuracies due to omitting higher order effects can be ruled out a priori. Additionally, the transport codes can be used as so-called black boxes and do not need to be adapted for computing sensitivity coefficients, typically by implementing adjoint flux computation capabilities.

A strong point of first order perturbation theory is that it can be combined with the Generalized Linear Least Squares (GLLS) method 
\citep{cecchini,humi,hemment,broadhead,saintjean,salvatores} to include measurements of integral variables in order to improve the knowledge about the nuclear data and, consequently, about the integral variables depending on them. 

The Monte Carlo approach, in contrast, has in the past been limited by the fact that an equally rigorous framework as the GLLS method was missing. However, this limitation has been removed recently by the development of the MOCABA methodology \citep{mocaba}. As for the GLLS method, MOCABA is based on a Bayesian model that permits the inclusion of information from integral measurements to improve the prediction of integral observables. A major advantage of this approach is that MOCABA updating can be applied directly to the integral observables without taking the detour via nuclear data updating. On the other hand, since nuclear data can be seen as just a special case of an integral observable,  MOCABA can also be used for non-perturbative nuclear data updating, which may be an attractive alternative to the GLLS approach for the generation of adjusted nuclear data libraries \citep{wpec_mocaba}. Mathematically, MOCABA can be seen as a non-perturbative generalization of the GLLS framework, which is shown by the fact that applying first order perturbation theory to the MOCABA model yields the well-known GLLS formulas \citep{mocaba}.

In this work, the performance of Monte Carlo-based Bayesian updating is tested for application in reactor physics. The question we address is what we can learn from previous reactor measurements for the prediction of future reactor cycles.  As a test case, we consider two consecutive burnup cycles of a Spanish  PWR nuclear power plant, denoted as Cycle~A and Cycle~B, where we attempt to predict Cycle~B based on measurements of Cycle~A. Here we focus on two reactor observables: the boron letdown curve, representing the critical boron concentration in the reactor as a function of burnup, and the reactor power distribution defined by the power values of the individual fuel assemblies in the reactor core. For the propagation of nuclear data uncertainties to the considered reactor observables, random samples of ENDF/B-VII.1 nuclear data files~\citep{endfb71} are generated with the aid of  the NUDUNA Monte Carlo code~\citep{nuduna}, which are then converted to random libraries for core simulations with the extensively validated SEANAP PWR analysis system~\citep{SEANAP2,SEANAP3}. The prior uncertainty distributions of the boron concentration and the assembly-wise power distribution are obtained from a statistical evaluation of the SEANAP results calculated with the different random libraries. Subsequently, the MOCABA updating procedure is used to improve the prior predictions of the reactor observables of Cycle~B by including measurements of Cycle~A. The predictive power of this approach is verified by comparing the predicted reactor observables of Cycle~B to the actual measured values. Finally, we compare the predictions from direct integral observable updating to those obtained via MOCABA updating of the SEANAP nuclear data library.

The paper is structured as follows. In Section~\ref{sect::descmethod}, we give a brief introduction to the SEANAP core analysis system, the NUDUNA nuclear data Monte Carlo code and the MOCABA updating scheme, and we describe how they are combined in our reactor uncertainty analysis. In Section~\ref{sect::applications} we apply this methodology to the PWR test case described above. After a discussion of the results, we end with conclusions and outlook.
%
%
\section{Description of the NUDUNA / MOCABA methodology applied to SEANAP core analysis} 
\label{sect::descmethod}
%
%
%
The procedure we use for reactor cycle prediction is divided in three steps:
\begin{itemize}
\item 
Generation of random nuclear data libraries with the aid of the NUDUNA Monte Carlo code,
\item
Performing SEANAP burnup simulations for Cycles A and B for each of the random library inputs,
\item
MOCABA updating of the prediction of Cycle~B using measurements of Cycle~A.
\end{itemize}
In the following, the codes involved in this procedure are described, together with the applied methodology.
%
%
%
\subsection{SEANAP: PWR core analysis}
\label{sect::SEANAP}
%
%
%
SEANAP is a system for the simulation and analysis of PWR cores, developed at Universidad Polit\'{e}cnica de Madrid \citep{SEANAP2,SEANAP3}. It consists of a chain of concatenated codes covering the different tasks needed to perform PWR core simulations, and it has been used for the simulation of many cycles of Spanish PWRs for the last 25 years, with good agreement between the simulated and the measured values. This system can calculate a wide variety of parameters: boron concentration, peak factors, axial offset, reactivity coefficients, power per fuel assembly, among many others. The SEANAP system is a fast simulation tool that allows the computation of many different observables at the core level. Hence, it is well suited to analyze the impact of uncertainties in the basic nuclear data on global core parameters.

\begin{figure}[t!]
  \begin{center}
    \includegraphics[width=0.45\textwidth]{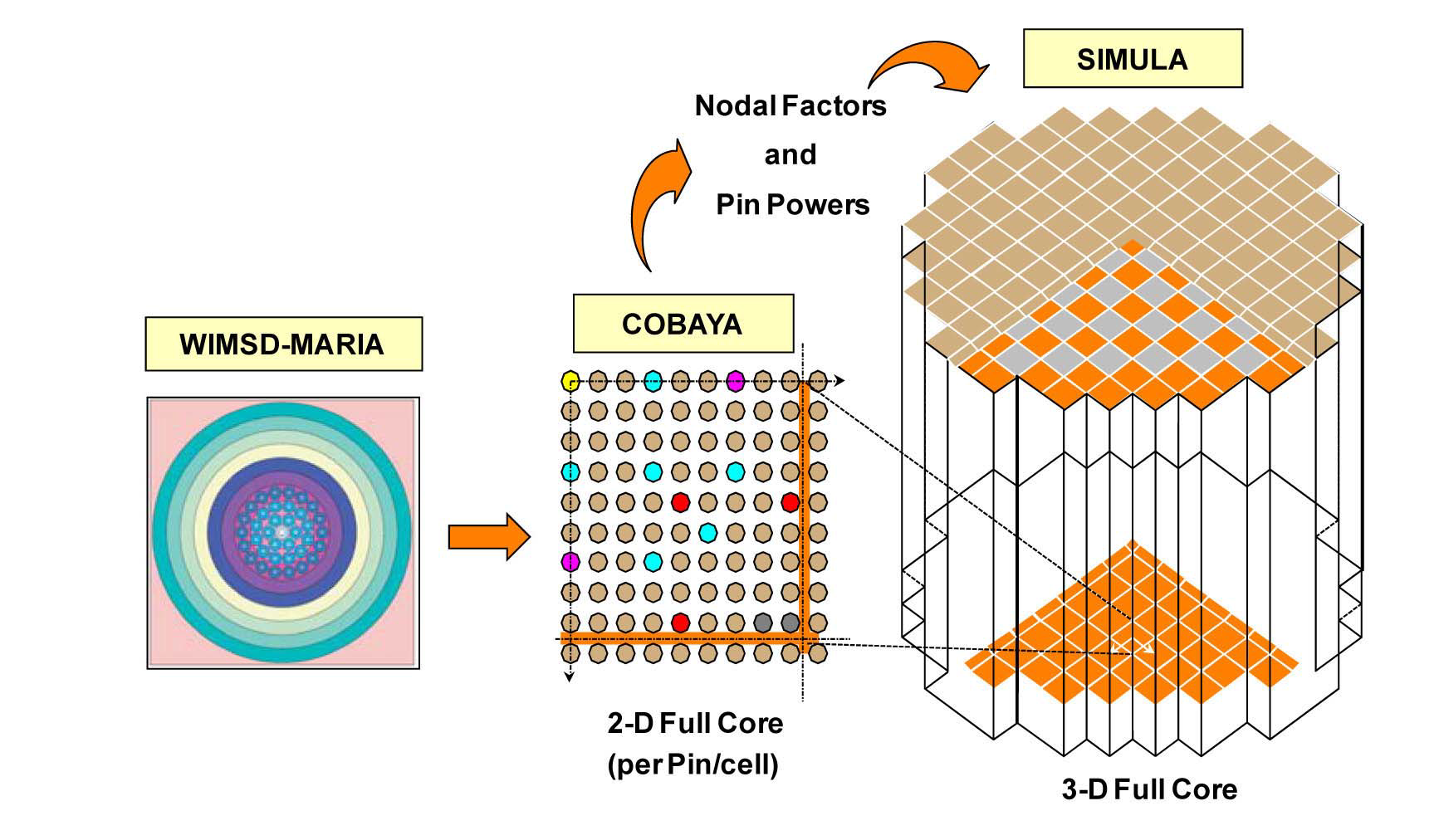}
    \caption[]{\label{fig::seanapfigure} Scheme of the PWR core analysis system SEANAP.}
  \end{center}
\end{figure}

The main components of this system are presented in Figure~\ref{fig::seanapfigure}. The MARIA subsystem consists of the PREWIM, WIMS-D4 and POSWIM codes. The PREWIM code generates the input files required by WIMS-D4 for all the fuel assemblies, covering the parameter space of the local physical variables, using a cylindricalized model of the square PWR assembly, which provides an efficient treatment of the PWR fuel assembly. The WIMS-D4 lattice code \citep{WIMS_traca,WIMSDinput} calculates the PWR fuel assembly in the annular cluster geometry by $S_N$ neutron transport calculation using a 69 energy group nuclear data library. As a result, the cross sections for each fuel assembly are obtained, homogenized at the assembly level and at the pin level.

COBAYA~2 is a 2D two-group pin-by-pin diffusion code, with a fine-mesh finite-difference method. Its purpose is to compute nodal discontinuity factors and hot-pin to node average power ratios, at some reference conditions, that are used in the SIMULA core simulator. SIMULA is a 3D nodal code, using four nodes per fuel assembly, an axial mesh of 34 nodes and simplified closed-channel thermal hydraulics. The solver is a linear-discontinuous finite-difference scheme for synthetic coarse-mesh few group diffusion calculation \citep{SEANAP1}. It uses the nodal discontinuity factors provided by COBAYA~2 for each node as a function of the burnup.

%
%
%
\subsection{NUDUNA: nuclear data random sampling}
\label{sect::NUDUNA}
%
%
%
The NUDUNA (NUclear Data UNcertainty Analysis) program, developed by AREVA GmbH \citep{nuduna}, provides Monte Carlo sampling 
of nuclear data, generating random libraries that can be used in core simulations.

This code reads evaluated nuclear data files in ENDF-6 format~\citep{endf6_format}, and provides random libraries based on the covariance information included in the evaluations for the following data:
\begin{itemize}
\item
average fission neutron multiplicities (File 1)
\item
resonance parameters (File 2)
\item
cross sections (File 3)
\item
angular distributions (File 4)
\item
decay data (File 8, Section 457)
\end{itemize}

NUDUNA is coupled to the NJOY nuclear data processing system \citep{njoy_manual} in order to automatically generate nuclear data inputs for different transport codes. For this work, the processing of the files to the WIMSD format has been implemented using the WIMSR module of NJOY, based on \citep{WIMSDupdate} and \citep{MacFarlaneProcessingEndf}. {NUDUNA also supports the ACE format of MCNP~\citep{mcnp5}, the AMPX format of  SCALE~\citep{scale60}, and the APOLLO~II neutron library format \citep{apollo2}.

Here we apply WIMSD-formatted NUDUNA samples in SEANAP transport calculations. The resulting statistics of reactor operation parameters reflects their uncertainty due to nuclear data uncertainties.

%
%
%
\subsection{MOCABA: Bayesian updating of predictions}
\label{sect::MOCABA}
%
%
%
MOCABA is a mathematical framework developed by AREVA GmbH to combine Monte Carlo sampling and Bayesian updating in order to achieve improved predictions of integral functions of nuclear data~\citep{mocaba}.
In this framework, the vector of application case variables $\mb y_A$, i.e.~the variables we want to predict, and the vector
of benchmark variables $\mb y_B$, for which measurements are available, are collected in a combined vector 
\begin{align}
  \mb y &\,=\, \left( \mb y_A^T  , \mb y_B^T \right)^T .
\label{eq::partition_y}
\end{align}
The fact that the application case and benchmark variables are known with limited precision is taken into account by treating $\mb y$ as a random vector, and the prior distribution $\rm p(\mb y)$, reflecting the uncertainty of $\mb y$ due to uncertainties of nuclear data, technological parameters and operational parameters, is assessed through Mon\-te Carlo sampling of these parameters and subsequent computation of $\mb y$ for each random sample. Then $\rm p(\mb y)$ is estimated from the resulting statistics of $\mb y$ computations. For a normal distribution model, this means to estimate the prior mean vector $\mb y_0$ and the prior covariance matrix $\bs\Sigma_0$ from the Monte Carlo data~\citep{mocaba}:\footnote{A more general class of distribution models can be accessed by making use of invertible variable transformations~\citep{mocaba}.}
\begin{align}
  \mb y_0 & \,=\, \left( \mb y_{0A}^T  , \mb y_{0B}^T \right)^T \,,\quad
 \bs\Sigma_0 \,=\,
\left(
 \begin{matrix}
 \bs\Sigma_{0A} & \bs\Sigma_{0AB} \\
 \bs\Sigma_{0AB}^T & \bs\Sigma_{0B}
 \end{matrix}
 \right) .
\label{eq::prior_mp}
\end{align}
Measurements of the benchmark vector $\mb y_B$ and/or linear constraints on $\mb y$, represented by a vector $\mb v$, are expressed in terms of a likelihood function ${\rm p}(\mb v \,|\, \mb y)$. To add this information to the prior information, Bayes' theorem is applied, which yields the posterior distribution:
\begin{align}
{\rm p}(\mb y \,|\, \mb v) & \,\propto\, {\rm p}(\mb v\, |\, \mb y) \, {\rm p}(\mb y)\,.
\label{eq::post_y}
\end{align}
For a normal distribution model, the Bayesian updating replaces the prior mean vector $\mb y_0$ and the prior covariance matrix $\bs\Sigma_0$ by the posterior mean vector $\mb y^*$ and the posterior covariance matrix $\bs\Sigma^*$:
\begin{align}
\mb y^* \,=\, \left( \mb y_A^{*T}  , \mb y_B^{*T} \right)^T\,,\quad
\bs\Sigma^* \,=\,
\left(
\begin{matrix}
\bs\Sigma_A^* & \bs\Sigma_{AB}^* \\
\bs\Sigma_{AB}^{*T} & \bs\Sigma_B^*
\end{matrix}
\right).
\label{eq::posterior_mp}
\end{align}
The impact of the benchmark measurements on the prediction of the application case variables is determined by the matrix $\bs\Sigma_{0AB}$ of prior covariances between application case and benchmark variables. If the corresponding correlations are close to 1, measurements of $\mb y_B$ have the potential to significantly improve the prediction of $\mb y_A$. This is generally the case if application case and benchmarks are similar, in the sense that they show similar responses to variations in the input parameters, in particular nuclear data. Due to its general formulation, the MOCABA framework can be applied to the prediction of any integral observable, and any integral measurement can be used to update the predictions. As mentioned above, MOCABA can also be used for non-perturbative updating of nuclear data libraries. For a more detailed description see \citep{mocaba}.
%
%
%
\subsection{Methodology of reactor cycle prediction}
\label{sect::methodology}
%
%
%
In the following, SEANAP, NUDUNA and MOCABA are used in combination to improve the predictions of reactor operation parameters in a given burnup cycle of a PWR (Cycle~B) based on measurements and simulations of the previous cycle (Cycle~A). 

\begin{figure}[ht!]
  \begin{center}
    \includegraphics[width=0.45\textwidth]{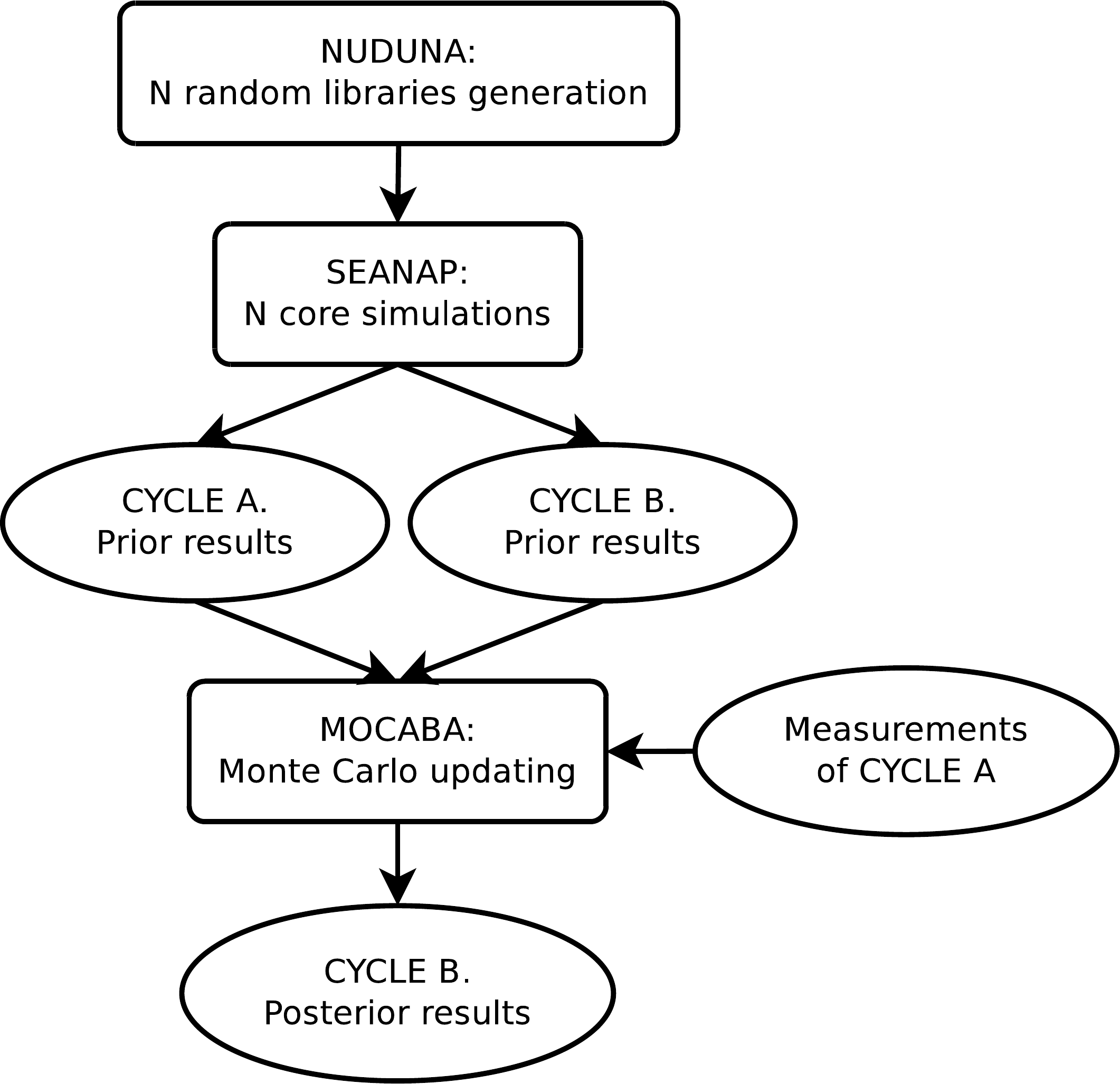}
    \caption[]{\label{fig::methodology} Scheme of the methodology.}
  \end{center}
\end{figure}

\Figure{}~\ref{fig::methodology} shows a scheme of the applied methodology, which is divided in three steps. 
First, a sufficient number (we use 200) of random nuclear data libraries are generated with NUDUNA, using covariance data from the ENDF/B-VII.1 evaluation \citep{endfb71} for the most important isotopes, being \textsuperscript{235}U, \textsuperscript{238}U, \textsuperscript{239}Pu, \textsuperscript{1}H, \textsuperscript{16}O and \textsuperscript{10}B. These random samples are combined with the reference library of WIMS using the  WILLIE code \citep{WILLIE}, yielding a set of random nuclear data libraries to be used in core simulations with SEANAP, which is the second step. The core simulation results are then used in a statistical analysis to estimate the prior mean values and covariances of the Cycle~A and Cycle~B observables.

In the last step, measurement information of Cycle~A is added by applying MOCABA updating to the prior Cycle~A and Cycle~B predictions. Hence, Cycle~A is used as a benchmark to improve the prediction of Cycle~B.
%
%
\section{Applications} 
\label{sect::applications}
%
%
%
Here we consider two different reactor observables: the boron letdown curve, defining the time-dependent critical boron concentration in the reactor coolant during a burnup cycle, and the fuel assembly-wise reactor power distribution. For the boron letdown-curve, the components of the integral observable vectors are the boron concentrations, and for the reactor power distribution they are the power values of the individual fuel assemblies in the reactor core, at different times during the burnup  cycle.

\begin{figure}[ht!]
  \begin{center}
    \includegraphics[width=0.3\textwidth]{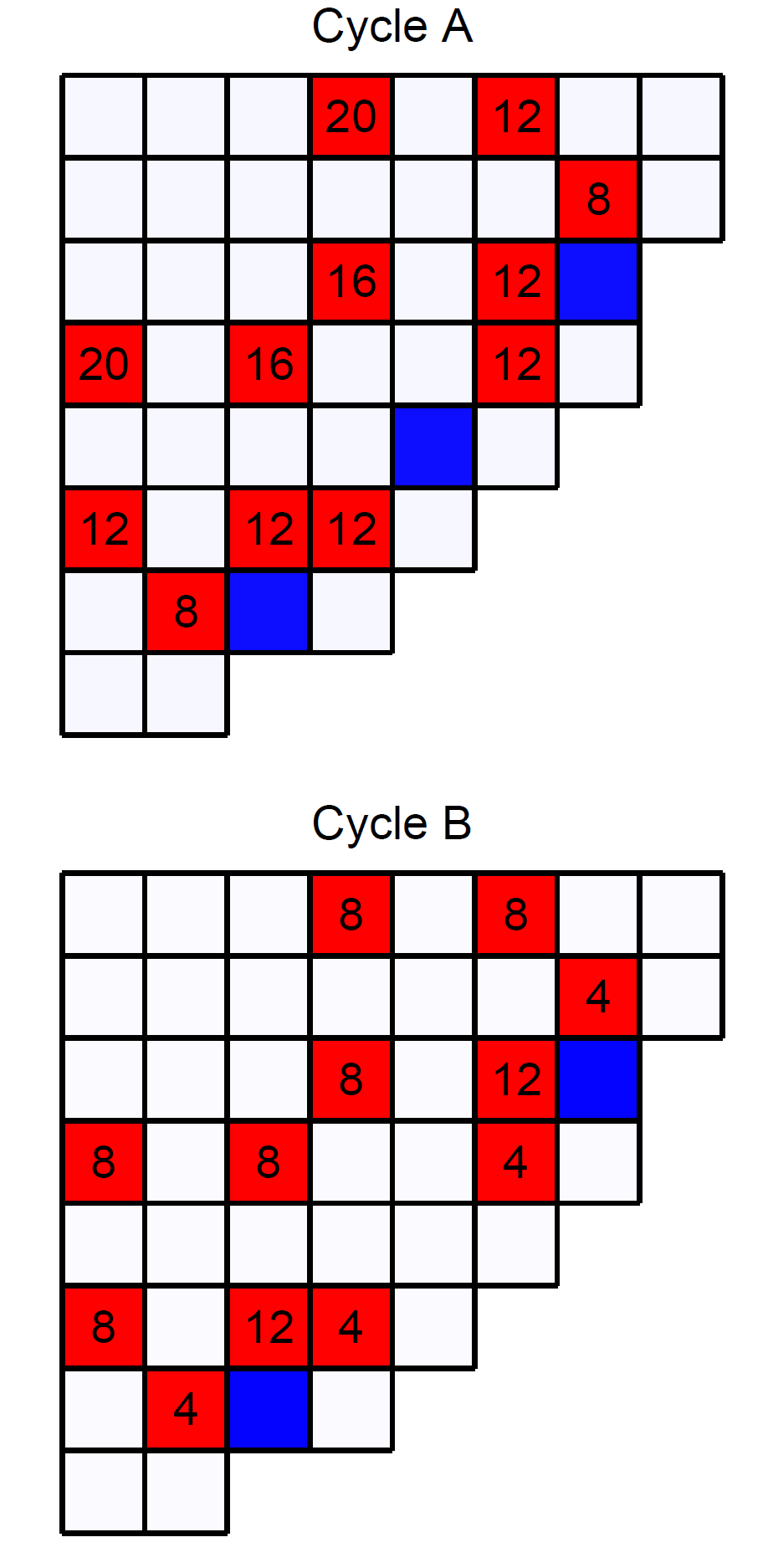}
    \caption[]{\label{fig::refueling_pattern} Refueling pattern for the two cycles In blue: fresh fuel without burnable absorber. In red: 
	fresh fuel with burnable absorber, the number indicates the number of rods with poison. In white: old fuel.}
  \end{center}
\end{figure}

The analyzed reactor contains 157 fuel assemblies, with a rated thermal power of 2775 MWt, using fresh and previously used fuel, and WABA rods as burnable absorber. The refueling pattern for both cycles is quite similar, although the number of burnable absorber rods in the fuel assemblies is higher in Cycle~A. The refueling patterns for both cycles are presented in \Figure{}~\ref{fig::refueling_pattern}, where all fresh fuel assemblies have an initial enrichment of 3.6~w/o, except for the fuel assemblies with 20 WABA rods in Cycle~A, whose enrichment is 3.24 w/o.

Measurements at 10 different burnup points of Cycle~A are used to update the Cycle~B predictions for 12 different burnup values. It is important to emphasize that only measurements of Cycle~A are used in the updating procedure, and comparisons of Cycle~B predictions to Cycle~B measurements are only made afterwards to assess the predictive power of the methodology. The uncertainty in the boron concentration measurements is 6~ppm,\footnote{The boron concentration values presented here refer to mass ratios of natural boron in the reactor coolant.} that is 0.6~\% at the beginning of the cycle (BOC); and the uncertainty in the power measurements is 5~\%.
%
%
%
%
\subsection{Boron letdown curve of a burnup cycle}
\label{sect::appBoron}
%
%
%
\begin{table*}[htb!]
  \begin{center}
    \caption{\label{tab::boronupd}Boron concentrations for Cycle~B as a function of burnup.}
    \footnotesize
    \begin{tabular}{ccccccc}
    \toprule
    Burnup (GWd/t) & $BC_{prior}$ & $\sigma_{prior}$ & $BC_{posterior}$ & $\sigma_{posterior}$ & $\sigma_{prior}/\sigma_{posterior}$ & $B_{measured}$\\
    \midrule
    0.13  & 986   & 46    & 986   & 4.2  & 11 & 983 \\
    1.34  & 867   & 45    & 868   & 3.3  & 13 & 874 \\  
    2.49  & 763   & 44    & 767   & 2.7  & 16 & 771 \\ 
    2.84  & 733   & 44    & 737   & 2.6  & 17 & 740 \\
    3.59  & 664   & 43    & 668   & 2.4  & 18 & 672 \\  
    4.44  & 590   & 43    & 595   & 2.3  & 18 & 596 \\ 
    5.55  & 487   & 42    & 494   & 2.2  & 19 & 496 \\
    6.69  & 383   & 42    & 391   & 2.4  & 18 & 394 \\  
    7.72  & 290   & 42    & 298   & 2.6  & 16 & 402 \\ 
    8.82  & 197   & 41    & 206   & 2.8  & 15 & 202 \\ 
    10.28 & 74    & 41    & 85    & 3.2  & 13 & 87 \\ 
    11.35 & -15   & 40    & -4    & 3.5  & 11 & 3 \\
    \bottomrule
    \end{tabular}
  \end{center}
\end{table*}

The evolution of the critical boron concentration during a reactor cycle is a measure of the cycle length. Hence, its prediction is essential for reactor operation planning. \Figure{}~\ref{fig::priorboron} shows the prior uncertainty in the predicted critical boron concentration of Cycle B due to nuclear data, obtained by application of the NUDUNA random libraries to the SEANAP calculations.

\begin{figure}[ht!]
  \begin{center}
    \includegraphics[width=0.45\textwidth]{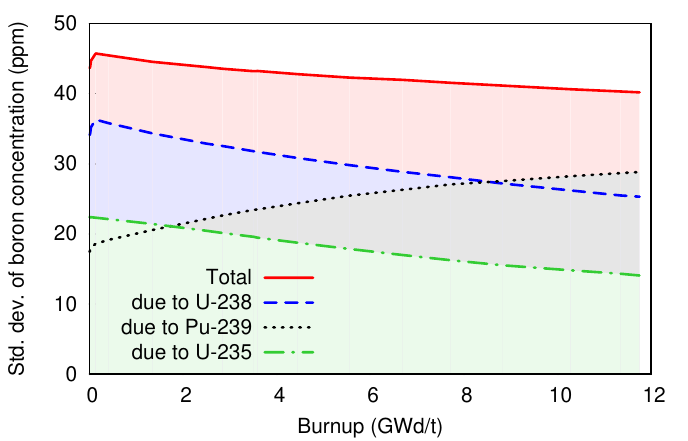}
    \caption[]{\label{fig::priorboron} Prior results for the standard deviation of the boron concentration for Cycle~B as a function of burnup.}
  \end{center}
\end{figure}

The total standard deviation due to all considered isotopes decreases slightly with burnup from a maximum value of 46~ppm at BOC to a value of 40~ppm at the end of the cycle (EOC).

\textsuperscript{238}U is  the principal uncertainty contributor for burnup values up to 8~GWd/t heavy metal. Its contribution decreases continuously from 36~ppm at BOC to 25~ppm at EOC. The uncertainties due to \textsuperscript{235}U and \textsuperscript{239}Pu have opposite behaviours. \textsuperscript{235}U is consumed during the cycle, so its importance is decreasing, with an effect of 
22~ppm at BOC and 14~ppm at EOC. On the other hand, \textsuperscript{239}Pu is formed from \textsuperscript{238}U during the cycle, so its uncertainty 
contribution increases from 19~ppm at BOC to 29~ppm at EOC.

The remaining materials have a much lower impact on the uncertainty of the boron concentration. Hydrogen and oxygen (in moderator and fuel) provide a basically constant uncertainty contribution of approx.~1~ppm and 5~ppm, respectively;  \textsuperscript{10}B causes 1~ppm uncertainty at BOC, and its impact is fully negligible at EOC.

A similar analysis of the prior uncertainties in the boron concentration was performed by \cite{SeanapUncert} with the Total Monte Carlo (TMC) code \citep{koning_tmc}, which was based on the TENDL-2012 nuclear data evaluation \citep{talys}. Comparing the results derived by \cite{SeanapUncert} to the results presented above, shows that uncertainties in the boron concentration are much larger for TENDL-2012/TMC than for ENDF/B-VII.1/NUDUNA, and also the relative importance of the isotopes differs. This confirms that the choice of a nuclear data evaluation may have a large impact on predictions and their corresponding uncertainties, as has already been demonstrated by \cite{CJDiezComparison}. 

\begin{figure}[ht!]
  \begin{center}
    \includegraphics[width=0.45\textwidth]{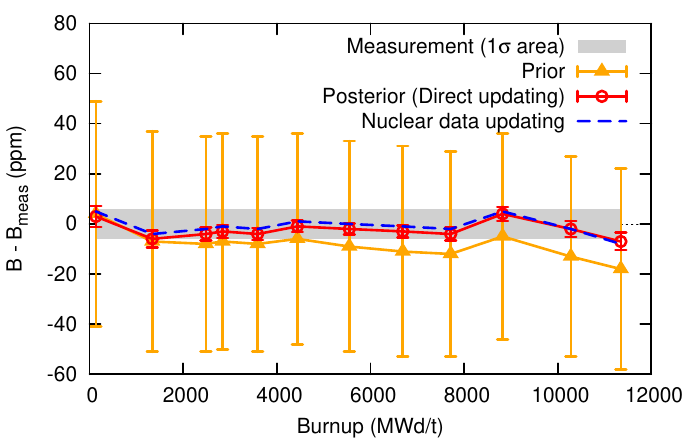}
    \caption[]{\label{fig::nucdatupd}Differences between measured boron concentrations and respective prior predictions, posterior predictions, and predictions obtained with a MOCABA-updated nuclear data library. The error bars represent the respective 1$\sigma$ uncertainties.}
  \end{center}
\end{figure}

Finally, Bayesian updating is applied to the prior boron letdown curve with the MOCABA procedure, taking into account measurements and simulations of the previous cycle, i.e.~Cycle A. The corresponding outcomes are shown in Table~\ref{tab::boronupd} and \Figure{}~\ref{fig::nucdatupd}. Two conclusions can be drawn from these results. First, for all burnup values, considering the previous cycle measurements leads to a better prediction of the boron letdown curve. Second, the uncertainty in the boron calculation is reduced by one order of magnitude, from values higher than 40~ppm to values around 3~ppm, which means a huge improvement for the prediction of a reactor cycle and, hence, for reactor operation planning.

The impressive improvement in the prediction of the boron letdown curve by including measurements of the previous cycle is explained by \Figure{}~\ref{fig::correlations}. It shows large correlations between the boron values of Cycle~A and Cycle~B in the range 0.92 - 0.99, which reflects a high similarity between application case and benchmark, i.e.~Cycle B and Cycle A.
\begin{figure}[ht!]
  \begin{center}
    \includegraphics[width=0.45\textwidth]{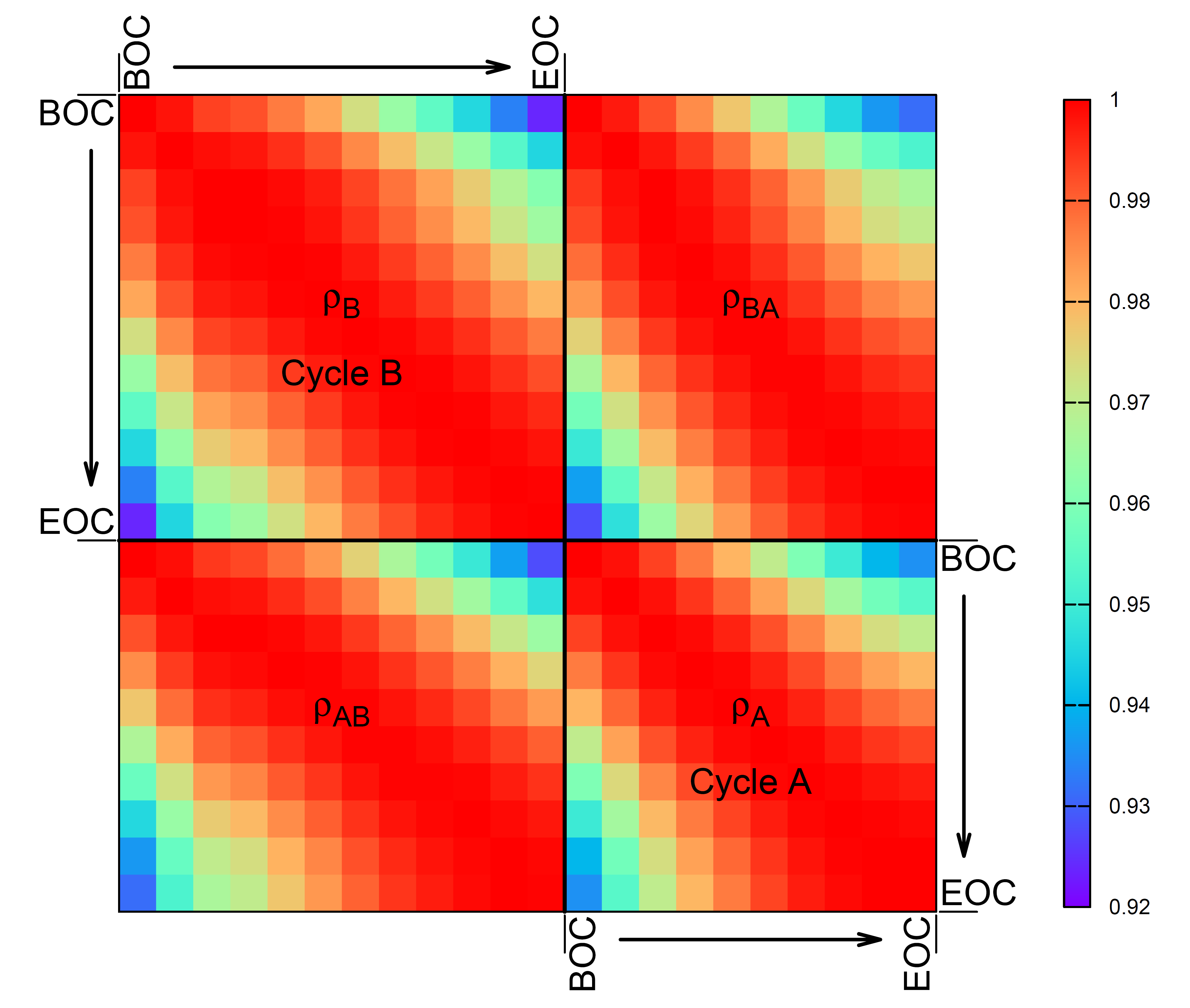}
    \caption[]{\label{fig::correlations} Correlation matrix of boron concentration values for Cycle~A and Cycle~B at different burnup steps.}
  \end{center}
\end{figure}

\Figure{}~\ref{fig::app7convergence} shows the convergence of the posterior boron concentration at a burnup of 7.72~GWd/t as a function of the number of benchmarks included in the Bayesian updating. The uncertainty is reduced with increasing number of benchmarks taken into account and thus the amount of information considered. Already including a single benchmark leads to a dramatic uncertainty reduction. Adding further measurements includes additional information, which leads to even better predictions. However, the simulation results for the different burnup steps are correlated, and thus the amount of new information of an additional measurement is limited. Hence, the impact of the first measurement is most prominent.

\begin{figure}[ht!]
  \begin{center}
    \includegraphics[width=0.45\textwidth]{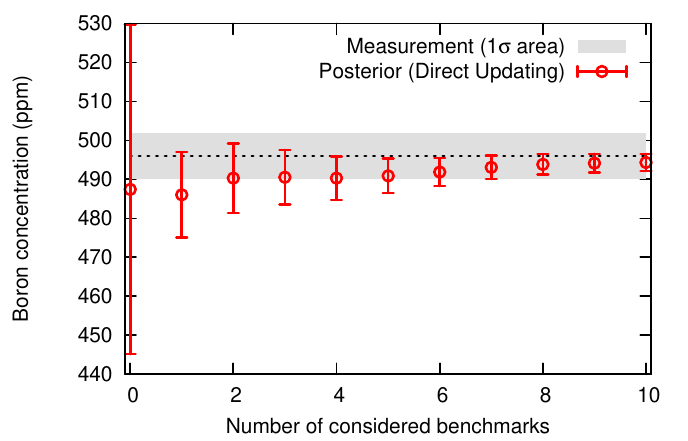}
    \caption[]{\label{fig::app7convergence}Convergence of the MOCABA updating procedure for increasing numbers of considered benchmarks for a burnup of 7.72~GWd/t.}
  \end{center}
\end{figure}

\subsection{Effects of nuclear data updating on boron letdown curve}
There is a different way of obtaining updated boron concentrations using MOCABA, which is updating the WIMS nuclear data library using the measurements of Cycle~A, and performing after that the SEANAP simulation for Cycle~B with the updated library as input. Here, we update the information of \textsuperscript{235}U, \textsuperscript{238}U, \textsuperscript{239}Pu and \textsuperscript{1}H in the 69 energy groups. In \Figure{}~\ref{fig::nucdatupd} the results obtained with direct updating of the boron letdown curve are compared to results obtained with the new nuclear data library. Both methods give consistent results which differ by less than 2~ppm.

Note that our study does not include other uncertainty contributions than nuclear data uncertainties, e.g., no uncertainties related to technological parameters or to neutronics and thermal hydraulic models are taken into account. This ignorance could result in adapting nuclear data in order to balance deficiencies in the description of other input values, which might result in unphysical updates of the nuclear data. To check this effect, we compare in \Figure{}~\ref{fig::Pu9upd} the differences between original and updated library and normalize the difference to the standard deviation of the original library. The largest deviations can be observed for the \textsuperscript{239}Pu fission cross sections, and the modifications are most prominent in the low-energy groups. However, the differences are always smaller than the standard deviation of the original values and can be explained by its uncertainties. Nevertheless, the update could compensate other deficiencies, and thus it is to be checked in future studies whether the updated library obtained for one specific reactor also improves the description of other reactors.

%
\begin{figure}[ht!]
  \begin{center}
    \includegraphics[width=0.45\textwidth]{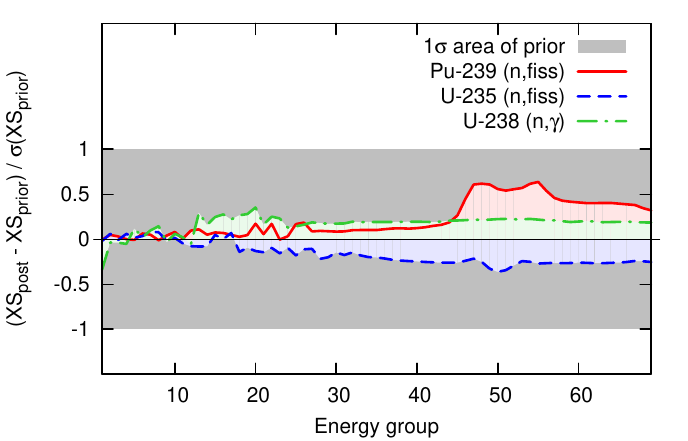}
    \caption[]{\label{fig::Pu9upd} Cross section updates normalized to one standard deviation.}
  \end{center}
\end{figure}
%
%
%
%
\subsection{Power per fuel assembly in a burnup cycle}
\label{sect::appPower}
%

%
%
\Figure{}~\ref{fig::powerBOCEOC} shows on the left-hand side NUDUNA/\linebreak SEANAP uncertainty estimates for the relative power per fuel assembly in Cycle~B at BOC and EOC. Here, the relative power values are the respective fuel assembly-wise power values divided by the mean power per fuel assembly in the reactor core. The uncertainty is largest in the center of the core and near the boundary, as has also been observed by \cite{santamarinapowermaps} and attributed to the radial swing of flux. 

\begin{figure}[]
  \begin{center}
    \includegraphics[width=0.45\textwidth]{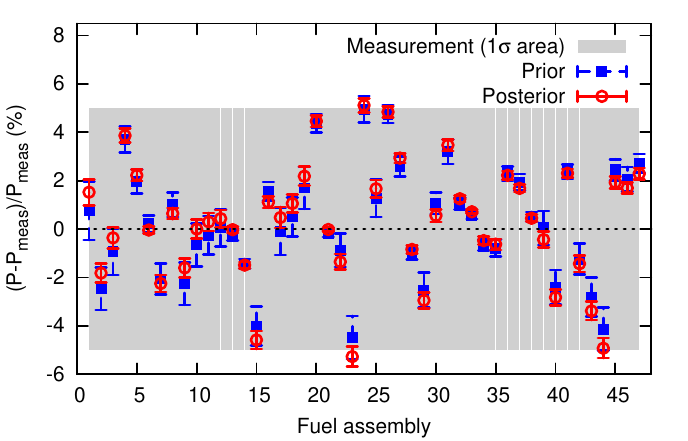}
    \caption[]{\label{fig::powermean2842} Relative deviations of the prior and posterior powers per fuel assembly of Cycle~B to its measurements at 2842~MWd/t.}
  \end{center}
\end{figure}

\begin{figure*}[]
  \begin{center}
    \includegraphics[width=0.7\textwidth]{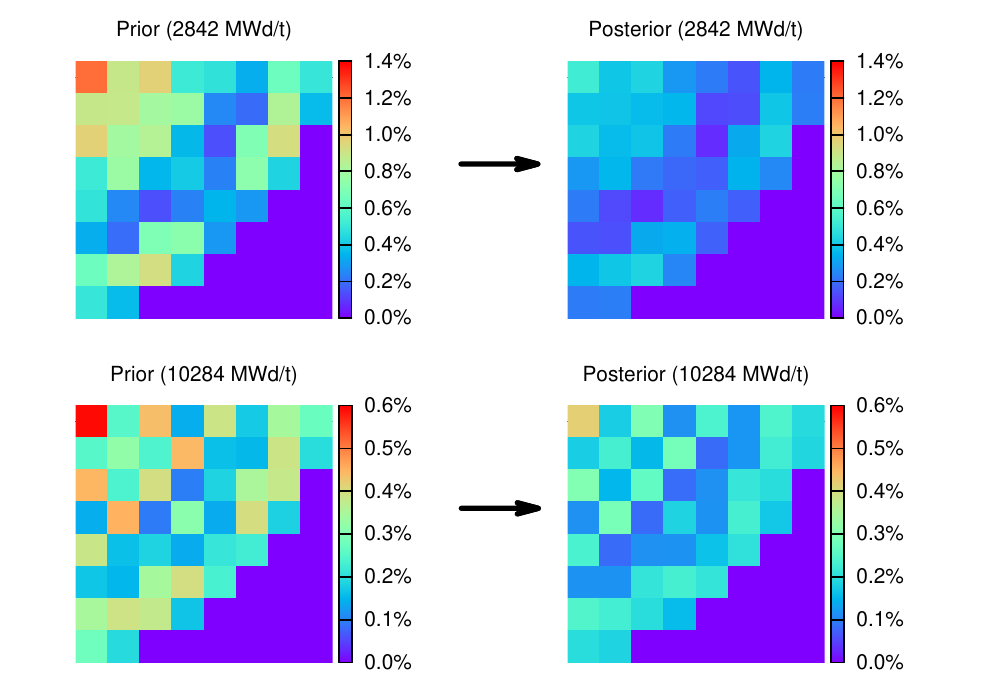}
    \caption[]{\label{fig::powerBOCEOC}Relative standard deviations for prior and posterior powers per fuel assembly for cycle burnups of 2842~MWd/t (top) and 10284~MWd/t (bottom) during Cycle~B.}
  \end{center}
\end{figure*}
The relative powers per fuel assembly have been determined during Cycle~A for each of the 47 fuel assembly positions in the considered quarter of the core and for 10 burnup steps, so there exist 470 power benchmark values. The right-hand graphs of \Figure{}~\ref{fig::powerBOCEOC} show the Cycle~B posterior uncertainties, which have been obtained with MOCABA based on the 470 Cycle~A benchmarks. The maximum uncertainty reduction amounts to approximately 50~\%. The average reduction depends on burnup (for 2842~MWd/t: 47~\%, for 10284~MWd/t: 29~\%), due to the fact that Cycle~A was shorter than Cycle~B and, consequently, the  correlations between Cycle~A and Cycle~B EOC simulations turn out to be less prominent. 

\Figure{}~\ref{fig::powermean2842} compares the simulation results to the Cycle~B measurements. However, due to the low precision of the measurements, it cannot be judged whether the posterior or prior estimates give a better description. Both are consistent with the measurements.

The analysis shows that the uncertainty reduction for the power predictions by MOCABA is much less favorable than for the boron letdown curve predictions. This feature results from the high uncertainties in the power measurements of 5~\%, which are far less precise than the boron measurements (6 ppm $\equiv$ 0.6~\%  uncertainty at BOC). The low precision of measurements can also not be balanced by their large quantity (470 power measurements instead of 10 boron concentration measurements). 

\subsection{Combining benchmarks and predictions of different responses}
With MOCABA, the prediction of a response can be improved by measurements of any benchmark observable. Thus it is also possible to update the power predictions by the boron concentration measurements and vice versa. In order to study this feature, we consider three scenarios for improving the power predictions:
\begin{itemize}
	\item Updating based on power measurements (see previous section),
	\item Updating based on boron concentration measurements,
	\item Updating based on all measurements.
\end{itemize}
\Figure{}~\ref{fig::powerWithAll} shows the results for a burnup of 2842~MWd/t during Cycle~B. If the power distributions are updated only with boron concentration measurements, then the uncertainty is reduced in maximum by 33~\% and on average by 13~\%. This rather modest improvement is explained by fairly low correlations of local power and boron concentration values. If the update considers both boron concentration and power measurements, then the uncertainty reduction amounts to on average 53~\% which is approx.~6~\% better than for considering only the power measurements. This demonstrates that different responses can be combined in the Bayesian updating, and all of them can provide useful information.

\begin{figure}[ht!]
  \begin{center}
    \includegraphics[width=0.45\textwidth]{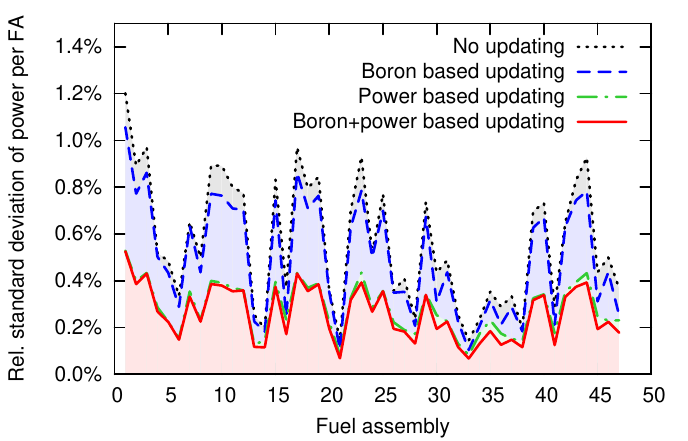}
    \caption[]{\label{fig::powerWithAll} Relative standard deviation of the power of each fuel assembly at 2842 MWd/t. Updates performed using only power per fuel assembly measurements, only boron concentration measurements, and both.}
  \end{center}
\end{figure}
}
%
%
\section{Conclusions} 
\label{sect::conclusions}
%
%
%
The nuclear data Monte Carlo code NUDUNA has been applied to PWR core simulations with SEANAP to provide both best estimates and their uncertainties for the boron letdown curve and the fuel assembly-wise power distribution. Next, the predictions for a given reactor cycle (Cycle~B) have been updated by applying the Bayesian inference model MOCABA, utilizing measurement information obtained in the preceding reactor cycle (Cycle~A). The resulting updated best estimates and their uncertainties have been compared afterwards to the measurements during Cycle~B in order to verify the predictive power of the procedure.

For the boron letdown curve, the MOCABA updating leads to major improvements for the best estimates and to massive uncertainty reductions. The nuclear-data-induced uncertainty of 40 to 45~ppm is reduced to 2 to 4~ppm after applying MOCABA, i.e.~to up to a 20 times lower uncertainty. 

For the power per fuel assembly, the 5~\% relative uncertainty of the power measurements of Cycle~A limits the possible reduction of the prediction uncertainty of Cycle~B. Still, MOCABA provides a major uncertainty reduction of 53~\% on average. This analysis also demonstrates the ability of the MOCABA methodology to combine simulations and measurements of different response variables.

MOCABA has also been applied to generate an updated nuclear data library in WIMS format based on the boron concentration measurements of Cycle~A. Computing the boron letdown curve for Cycle~B with this updated library gives results that are almost identical to the ones obtained by direct MOCABA updating of the boron concentrations (differences are smaller than 2 ppm). The nuclear data in the resulting library lie within the one-standard-deviation range of the original data, and thus are compatible with them. However, the updating might also compensate for calculation code deficiencies or technological parameter uncertainties of the considered PWR plant. Future studies need to address the question whether the updated nuclear data obtained for one specific reactor with one specific reactor code suite can improve the description of other analyzes.

The blind tests presented in this paper show very good performance of the NUDUNA/MOCABA best estimate plus uncertainty methodology. The obtained gain in precision, especially for the boron letdown curve, is impressive and promises major economical benefits. Compared to the traditional perturbative GLLS method, our Monte Carlo-Bayes procedure has four major advantages: it can be easily implemented since the transport codes used for the computation of the integral observables of interest can be treated as black boxes; it can easily address any integral function of nuclear data, such as boron concentration, power per fuel assembly, axial offset, and peak factors; integral observables can be directly updated without taking the detour via nuclear data updating, although also updated nuclear data libraries can be generated; and, due to its non-perturbative nature, it is not limited to the regime where nuclear data uncertainties are small.
%
%
%
%
%

\section*{Acknowledgements}
This work was supported by AREVA GmbH and conducted in the framework of the agreement in the area of Propagation of Uncertainties for Neutronic Calculations in Criticality Safety Analysis between the Spanish Nuclear Safety Council (CSN) and Universidad Polit\'ecnica de Madrid (UPM).


\section*{References}
\bibliographystyle{elsarticle-harv}
\bibliography{nuduna_mocaba_seanap}

\end{document}